\documentclass[ reprint, amsmath,showpacs,amssymb,aps]{revtex4-1}
 
\pdfoutput=1
\usepackage{epsfig,graphicx,xcolor,amsbsy,amssymb,latexsym,amsfonts,amsmath}
\usepackage{pstricks}
\usepackage{color}

\usepackage{eurosym}
\usepackage{graphicx}
\usepackage{tikz}
\usepackage{amsmath,amssymb,amsfonts,pstricks}
\newcommand{\bea}{\begin{eqnarray}\displaystyle}
\newcommand{\eea}{\end{eqnarray}}

\vspace{-0.5cm}
\begin{document}
\title{
%\begin{flushright}{\vspace{-2.5cm}\small SNUST 15-07\\}\end{flushright}
\vspace{-1.8cm}
\bf{Quantum Tunneling Of Electron Snake States In An Inhomogeneous Magnetic Field}\\[15pt]}
\author{Pervez Hoodbhoy}
%\date{\today}
\affiliation{Department of Physics\\
Forman Christian College\\
Lahore, Pakistan.}

\begin{abstract}
	In a two dimensional free electron gas (2DEG) subjected to a perpendicular spatially varying magnetic field, the classical paths of electrons are snake-like trajectories that weave along the line where the field crosses zero. But quantum mechanically this system is described by a symmetric double well potential which, for low excitations, leads to very different electron behavior. We compute the spectrum, as well as the wavefunctions, for states of definite parity in the limit of nearly degenerate states, i.e. for electrons sufficiently far from the $B_z=0$ line. Transitions between the states are shown to give rise to a tunneling current. If the well is made asymmetrical by a time-dependent parity breaking perturbation then Rabi-like oscillations between parity states occur. Resonances can be excited and used to stimulate the transfer of electrons from one side of the potential barrier to the other through quantum tunneling.
	\bigskip
	\\
	\bf{To be published in Journal of Physics:Condensed Matter}

\end{abstract}
\maketitle

Snake orbitals exist in mesoscopic systems where electrons may be presumed
to move freely in a two-dimensional electron gas (2-DEG) such as created by
using a GaAs/GaAlAs heterostructure. If such a system is subjected to a
perpendicular magnetic field whose strength varies linearly from one edge of
the sample to the other then, classically speaking, an electron moving close
to the line where $B_{z}=0$ alternately experiences regions of oppositely
directed fields and so the Lorentz force redirects trajectories towards this
boundary. In what becomes effectively a 1-D magnetic quantum wire, electrons
can propagate unidirectionally in snake-like fashion. The first quantum
calculation of electron trajectories and the energy spectrum of such a
system was carried out by Muller \cite{Muller}. This involved numerical
diagonalization of the Hamiltonian in a harmonic oscillator basis. Using
qualitative reasoning, he pointed out that snake states weaving around the $
B_{z}=0$ boundary break time reversal symmetry in the sense that electrons
close to the boundary can propagate as free particles in one direction but
not in the other. Following up on Muller's calculation, Reijniers and
coworkers \cite{Reijners1} \cite{Reijners2} performed extensive numerical
calculations and were able to give a clearer picture of such states,
including their confinement due to interference between oppositely directed
waves. However those calculations were done in a model that is somewhat
unrealistic, i.e. one in which the magnetic field increases by a discrete
step.

Calculations in inhomogeneous fields assume relevance because studying
electron motion in appropriate mesoscopic situations has been well within
experimental grasp for many years now. Nogaret \cite{Nogaret} has
reviewed electron dynamics in inhomogeneous fields and experimental strategies for making such fields. Recent technological advances have allowed the fabrication of high mobility 2-DEGs with nanomagnets of well defined shapes placed above or below, allowing one to study the effect of inhomogeneous $B$ fields with different profiles. These distributions can profoundly influence electron transport properties and may lead towards the design of useful magnetoelectronic devices. Using the idea of classical snake states one may compute, for example, the magnetoresistance of such systems \cite{Rectification} - only to discover that this picture of electron motion becomes progressively incorrect as $B$ is increased and the electron density decreased. Quantum effects become more pronounced in these situations. This phenomenon has been observed by Schuler et. al \cite{Expt2} who studied resonant transmission through electronic quantum states that exist at the zero points inside a ballistic quantum wire. They explored the dependence on the amplitude of the magnetic field as well as on the Fermi energy. Maha et. al \cite{Expt1} have observed asymmetric transport and rectification effects associated with snake states.

The present paper examines the nature of snake states in a region where
quantum properties are dominant and semi-classical arguments become
inapplicable. The goal is to make as much progress analytically as possible;
while numerical calculations are often essential, one pays the price of
decreased intuition. In the case under consideration, although one cannot
fully succeed in achieving closed form solutions, some progress can be made
by invoking ideas familiar from WKB theory and instanton methods.

The specific situation considered here is as in ref \cite{Muller}, i.e. a
magnetic field is directed perpendicularly to a 2-DEG that varies linearly
in strength from one edge of the sample to the other. Electrons in Landau
levels on opposite sides of the $B_{z}=0$ boundary have identical energies
if the Zeeman coupling is ignored. This degeneracy results in tunneling
across a potential barrier. The consequence is a current transverse to the
field gradient. States of opposite parity are shown to carry currents in
opposite directions. We show that if an additional external time-varying
electric field of appropriate frequency is imposed, Rabi-like oscillations
occur and these can be used to identify the orbital centers of the tunneling
electrons. This mechanism allows electrons in Landau orbits to swap
positions across the  $B_{z}=0$ line.

\section{Preliminaries}

Our starting point is the 2-D Schr\"{o}dinger equation describing free
electrons confined to a rectangular $(L_{x},L_{y})$ sample. A $\hat{z}$%
-directed magnetic field increases linearly with $y$, $\ \vec{B}%
(y)=(0,0,yB^{\prime })$.\ The origin of coordinates will be taken to be at
the sample's centre. The Hamiltonian is, 
\begin{equation}
H=\frac{1}{2m}\left( \mathbf{p}-\frac{e}{c}\mathbf{A}\right) ^{2}.
\label{Ham}
\end{equation}
The gauge potential is chosen as, 
\begin{equation}
\mathbf{A=}-\hat{x}\frac{1}{2}y^{2}B^{\prime }.
\end{equation}%
This potential is translationally invariant in $x$. This implies a plane wave solution, $\psi (x,y)=\frac{1}{\sqrt{L_{x}}}e^{-ik_{x}x}\varphi (y)$. It is convenient to define a relevant magnetic
length scale $L_{M},$

\begin{equation}
L_{M}=\left( \frac{2\hslash c}{eB^{\prime }}\right) ^{1/3}.
\end{equation}%
After rescaling, $\varphi $ satisfies a reduced
eigenvalue equation with the single parameter $\alpha ,$ 
\begin{equation}
\left( -\frac{1}{2}\frac{d^{2}}{d\eta ^{2}}+V(\eta )\right) \varphi
=\epsilon \varphi \,,\text{ }V(\eta )=\frac{1}{2}(\eta ^{2}-\alpha
^{2})^{2}\,.  \label{SE1}
\end{equation}%
In the above, 
\begin{equation}
\eta =\frac{y}{L_{M}},\text{ \ }\alpha ^{2}=k_{x}L_{M},\text{ \ }E=\frac{
	\hslash ^{2}}{mL_{M}^{2}}\epsilon .
\end{equation}%
The quantity $k_{x}$ is the eigenvalue of $\hat{p}_{x}=$ $\frac{\hslash }{i}%
\frac{\partial }{\partial x}$. One notes that $\hat{p}_{x}$ is not the
canonical momentum operator and so $k_{x}$ is unrelated to the physical
momentum. Translational invariance in $x$ allows periodic boundary
conditions $\psi (x+L_{x},y)=\psi (x,y)$ to be imposed. The sum over $k_{x}$
(which, of course, can take either sign) can therefore be converted to an
integral, 
\begin{equation}
\ \sum_{k_{x}}\longrightarrow \frac{L_{x}}{2\pi }\int dk_{x}\ =\frac{1}{\pi }%
\frac{L_{x}}{L_{M}}\int \alpha d\alpha .  \label{dens}
\end{equation}%
The sum over $k_{x}$ means that we shall have to deal with values of $\alpha
^{2}$ that can take both positive and negative values. In the case $k_{x}>0$
there are minima of the potential located at $\eta =\pm \alpha $. On the
other hand, for $k_{x}<0$ the two minima coalesce at $\eta =0.$\ \ The
eigenfunctions for the two cases also, of course, have qualitatively
different behaviors.

The double well oscillator of Eq.\ref{SE1} is famously unsolvable and has no exact solutions. But this seemingly simple potential actually has rich structure and has been pursued because it occurs in quantum field theory as well as various branches of physics and molecular chemistry. In fact it has been the subject of nearly a thousand investigations, many of which have followed the pioneering work of Bender and Wu \cite{Bender1}-\cite{Bender2}. Of course, one can always seek
approximate solutions. In the limit of large negative $\alpha ^{2}$ the eigenvalues are easily found to be, 
\begin{equation}
\epsilon _{n}=\frac{1}{2}\alpha ^{4}+\left( n+\frac{1}{2}\right) \sqrt{%
	2\left\vert \alpha ^{2}\right\vert },  \label{left}
\end{equation}%
or, upon restoring units,%
\begin{equation*}
E_{n}=\frac{\hslash ^{2}k_{x}^{2}}{2m}+\left( n+\frac{1}{2}\right) \sqrt{%
	\frac{e\hslash ^{3}B^{\prime }}{m^{2}c}\left\vert k_{x}\right\vert },
\end{equation*}

The lowest eigenvalues of Eq.\ref{SE1} are also easily obtained in the limit
of large positive $\alpha ^{2}.$ These correspond to the two minima being
widely separated although, for now, we do not know what this means in
quantitative terms. Intuitively it should mean the absence of
appreciable tunneling between the well minima because of the intervening
high potential barrier whose maximum is $V(0)=\frac{1}{2}\alpha ^{4}$. In
this situation the double well potential can be harmonically approximated by
two single isolated wells,\ $V(\eta )=\frac{1}{2}(2\alpha )^{2}(\eta \pm
\alpha )^{2}$ with $2\alpha $\ being the classical oscillation frequency.
The normalized eigenfunctions are the usual Hermite polynomials, 
\begin{equation}
\frac{1}{\sqrt{2^{n}n!}}\left( \frac{2\alpha }{\pi }\right) ^{1/4}H_{n}[%
\sqrt{2\alpha }(\eta \pm \alpha )]e^{-\alpha (\eta \pm \alpha )^{2}},
\label{Herm}
\end{equation}%
that correspond to the energy, 
\begin{equation}
\epsilon _{n,\alpha ,\pm }=2\alpha \left( n+\frac{1}{2}\right) .  \label{sho}
\end{equation}%
In this extreme asymptotic limit the eigenstates are degenerate and may be
chosen to have definite parity, 
\begin{equation}
\varphi _{n}^{\pm }(\eta )=\frac{1}{\sqrt{2}}\left[ \varphi _{n}(\eta
-\alpha )\pm \varphi _{n}(\eta +\alpha )\right] .
\end{equation}%
Note that the energy of a gyrating electron increases linearly with $\alpha ,
$ i.e. in either direction from the sample's centre. This is in contrast to
the case of a uniform field where the energy is flat and Landau levels of a
given $n$ are massively degenerate.

By expanding in a harmonic oscillator basis, one can numerically diagonalize
the Hamiltonian in Eq.\ref{SE1}. In Fig.1 the lowest few eigenvalues are
plotted as a function of $\alpha ^{2}$ for both signs of $\alpha ^{2}$. One
can see the emergence of a band structure. The asymptotic conditions leading to Eq.\ref {left} and Eq.\ref{sho} \ are well satisfied. The confining walls can be chosen far enough away so as to have negligible influence upon the
wavefunctions on the $y=0$ line.
\begin{figure}[h]
	\includegraphics[width=9cm, height=7cm]{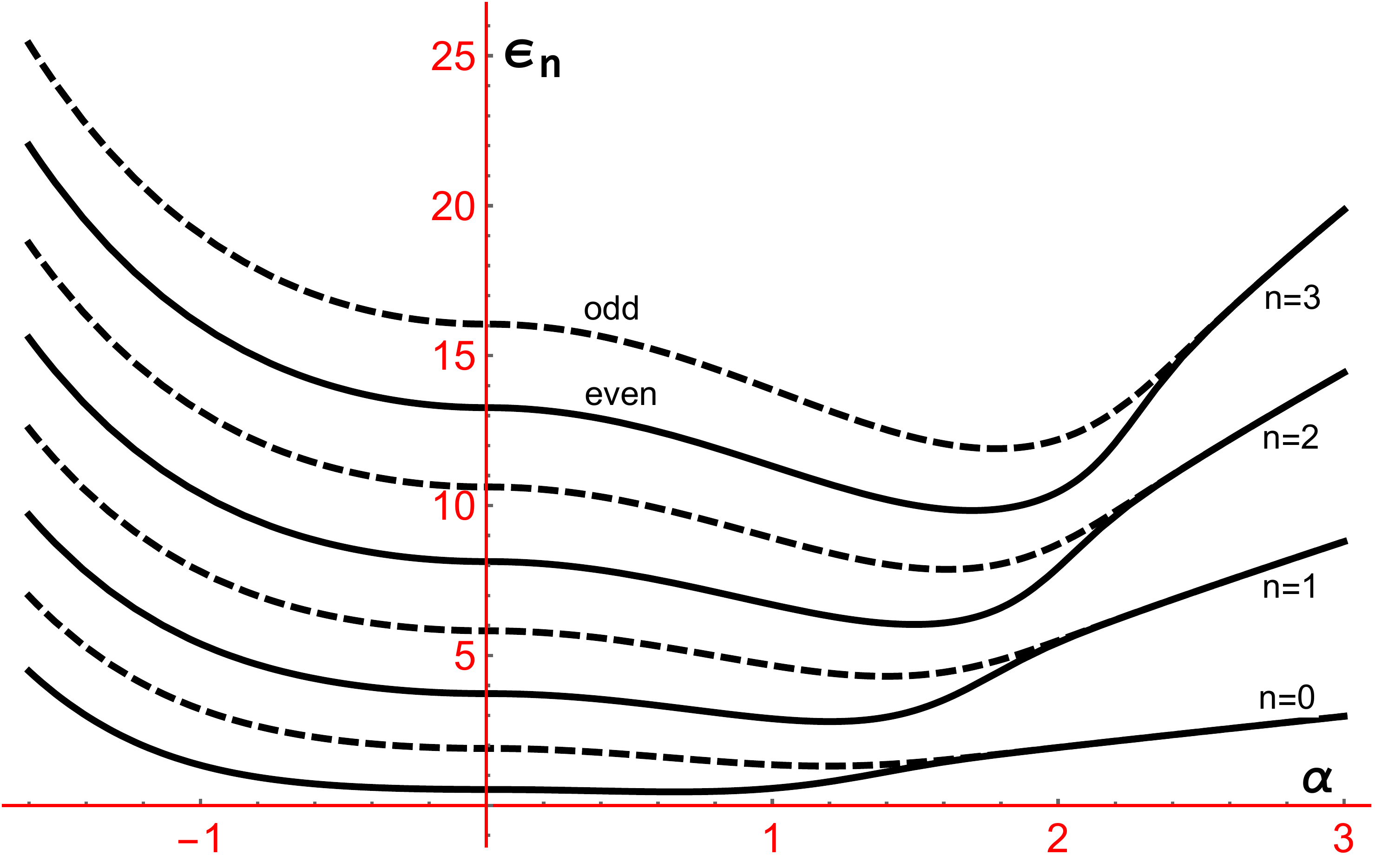} \centering
	\caption{Energies of the lowest electron states obtained through numerical diagonalization. Lower/upper curves for a given $n$ are of even/odd parity. These are plotted as a function of $\protect\alpha ,$\ with $\alpha ^{2}>0$ towards the right and  $\alpha ^{2}<0$ on the left. }
\end{figure}
\bigskip

Consider now the interval of $\alpha $\ where an electron is almost, but not wholly, confined to the two well minima. If one is originally centred at $\eta =\alpha$, it will bounce within the confines of its own well but will occasionally manage to tunnel across to the other well at $\eta =-\alpha $\ before returning after some finite time. For larger $\alpha ^{\prime }s,$ i.e larger separation between the two minima, tunneling will become less frequent. Note that $\alpha ^{-1}$ plays the role of $\hslash $\ in the usual quantum double oscillator, the smallness of which is crucial to the success of semi-classical methods. These methods are essential because non-analyticity in the coupling constant makes perturbation theory useless. Over the years, in a tour de force, Zinn-Justin and collaborators \cite{Zinn1}-\cite{Zinn2} have used sophisticated resurgent trans-series techniques involving a leading-order summation of multi-instanton contributions to the path integral representation of the partition function for calculating corrections to the leading order result for eigenvalues.  
After appropriately scaling their results to the present case, the first few
terms for $n=0$ and $n=1$ are: 
\begin{eqnarray}
\epsilon _{0,\pm } &=&\alpha \mp \frac{1}{\sqrt{\pi }}2^{\frac{5}{2}}\alpha
^{\frac{5}{2}}e^{-\frac{4\alpha ^{3}}{3}}  \nonumber \\
&&\times \left( 1-\frac{71}{96\alpha ^{3}}-\frac{6299}{18432\alpha ^{6}}-%
\frac{2691107}{5308416\alpha ^{9}}\right)   \label{m1} \\
\epsilon _{1,\pm } &=&3\alpha \mp \frac{1}{\sqrt{\pi }}2^{\frac{13}{2}%
}\alpha ^{\frac{11}{2}}e^{-\frac{4\alpha ^{3}}{3}} \nonumber \\
&&\times \left( 1-\frac{347}{96\alpha ^{3}}+\frac{5317}{18432\alpha ^{6}}-%
\frac{15995159}{5308416\alpha ^{9}}\right)   \label{m2}
\end{eqnarray}

To be added to the above are parity independent terms whose first 300, 70
terms for $n=0,1$ respectively have been calculated by the authors in
perturbation theory. These shall not be of interest here, but we shall use
in Fig.2 the correction factors in the brackets above. \bigskip

\section{Wavefunctions}

We now consider electrons orbiting around gyration centres located at $\pm
\alpha $. The Hamiltonian Eq.\ref{SE1} is even under $\eta \rightarrow -\eta 
$ so one can restrict attention to the region $\eta \geq 0$. We follow the
philosophy of Coleman\cite{Coleman} by insisting that the solution inside a
well centred at $\alpha $\ be well approximated by a wavefunction close to
that of a SHO. For energies sufficiently below the well maximum $\alpha
^{4}/2$, tunneling splits the ground state's degeneracy by a small amount
proportional to $\delta $ with $\delta >0.$\ The positive parity state, by
virtue of having smaller curvature (less kinetic energy), has lower energy, 
\begin{equation}
\epsilon _{n\pm }=\left( n+\frac{1}{2}\right) 2\alpha (1\mp \delta ),\text{
	\ }\alpha ^{3}>>2\left( 2n+1\right) .\text{\ }
\end{equation}%
Here $n$\ is a non-negative integer. Near the well minimum $\eta =\alpha $
the potential is quadratic and for the positive parity state Eq.\ref{SE1}
becomes,%
\begin{equation}
-\frac{1}{2}\frac{d^{2}\Phi _{\pm }}{d\eta ^{2}}+\frac{1}{2}4\alpha
^{2}(\eta -\alpha )^{2}\Phi _{\pm }=\epsilon _{n\pm }\Phi _{\pm }\,.
\label{q1}
\end{equation}%
\qquad To take the analysis further, let us define a new variable $z$ and
rescaled energy $\epsilon $, 
\begin{eqnarray*}
	z &=&2\sqrt{\alpha }(\eta -\alpha ) \\
	\varepsilon _{n\pm } &=&\frac{\epsilon _{n\pm }}{2\alpha }.
\end{eqnarray*}%
Temporarily suppressing the indices for clarity, this allows Eq.\ref{q1} to
be more simply written as,

\begin{equation}
\left( -\frac{d^{2}}{dz^{2}}+\frac{1}{4}z^{2}-\varepsilon \right) \Phi =0\,,
\label{q2}
\end{equation}%
the solutions of which are the parabolic cylinder functions,%
\begin{equation}
\Phi (z)=c_{1}D_{\varepsilon -\frac{1}{2}}(z)+c_{2}D_{-\varepsilon -\frac{1}{
		2}}(iz).
\end{equation}%
We first consider solutions for $z>0$. If the confining walls of the sample
are sufficiently far away, then the second term must be discarded because
for large positive $z$ this goes as $e^{z^{2}/4}$ and so we must choose $%
c_{2}=0$. Thus the wavefunction around the well minimum is, 
\begin{equation}
\Phi (z)=N_{\varepsilon }D_{\varepsilon -\frac{1}{2}}(z),  \label{Right}
\end{equation}%
where $N_{\varepsilon }$ is a normalization factor. Recalling that $%
\varepsilon _{n\pm }=(n+\frac{1}{2})(1\mp \delta )$ where $n$ is an integer,
we recall that for $\delta =0$\ the parabolic cylinder functions can be
expressed in terms of the Hermite polynomials, 
\begin{equation*}
D_{n}(z)=2^{-\frac{n}{2}}H_{n}(z/\sqrt{2})e^{-z^{2}/4}.
\end{equation*}

To the left of the well minimum (i.e. $\eta <\alpha $ ) $D_{\varepsilon - 
	\frac{1}{2}}\left(z\right) $ blows up as $z\rightarrow -\infty $. However
the domain is actually finite, i.e. $z>-2\alpha ^{3/2}$. To leading order in 
$\delta$ the asymptotic behavior in this region is,

\begin{equation}
\Phi _{n\pm }(z)=N_{n\pm }\left( z^{n}e^{-\frac{z^{2}}{4}}+\frac{\left(
	-1\right) ^{n+1}\sqrt{2\pi }}{\Gamma \lbrack -n\pm (n+\frac{1}{2})\delta ]}%
\frac{e^{\frac{z^{2}}{4}}}{z^{n+1}}\right) .  \label{phi}
\end{equation}

\bigskip

For small $\delta $ the gamma function is 
\begin{equation*}
\Gamma \lbrack -n\pm (n+\frac{1}{2})\delta ]=\pm \frac{\left( -1\right) ^{n}%
}{n!}\frac{1}{(n+\frac{1}{2})\delta }+O(\delta ^{0}),
\end{equation*}%
and therefore the wavefunction for widely separated wells and fixed negative 
$z$ is, to leading order,

\begin{equation}
\Phi _{\pm }(z)=N_{n\pm }\left( z^{n}e^{-\frac{z^{2}}{4}}\mp \delta \sqrt{
	2\pi }(n+\frac{1}{2})n!\frac{e^{\frac{z^{2}}{4}}}{z^{n+1}}\right) ,\text{ }\
z<0.  \label{Left}
\end{equation}

\bigskip

We now have the parts of the wavefunction to the right and left of the well
minimum. However Eq.\ref{Left} evidently does not have the correct parity
behavior under $\eta$ $\rightarrow$ $-$$\eta $, i.e. these approximate
solutions do not obey $\varphi _{+}^{\prime }(0)$\ $=0$\ and $\varphi
_{-}(0) $\ $=0$. Hence we must return to Eq.\ref{SE1} and construct
appropriate solutions near $\eta =0.$ We therefore look for solutions in the
presence of the full quartic potential (i.e. solve Eq.\ref{SE1} rather than
just the quadratic approximation to it). Inspired by the WKB method but
independent of it, let us make the ansatz: 
\begin{equation}
h(\eta )e^{-\alpha ^{2}\eta +\frac{1}{3}\eta ^{3}},
\end{equation}%
with $h(\eta )$ a function to be determined. Substituting into Eq.\ref{SE1}
yields, to leading order in $\delta $, \ 
\begin{equation}
h^{\prime \prime }-2(\alpha ^{2}-\eta ^{2})h^{\prime }+2(\alpha +2\alpha
n+\eta )h=0.
\end{equation}%
For large $\alpha $ we can ignore the first term and obtain a first order
equation that is immediately solvable,

\begin{equation*}
h(\eta )=\frac{(\eta +\alpha )^{n}}{(\eta -\alpha )^{n+1}}
\end{equation*}%
Since Eq.\ref{SE1} is even under $\eta \longleftrightarrow -\eta $ it
follows that $h(-\eta )e^{\alpha ^{2}\eta -\frac{1}{3}\eta ^{3}}$ is equally
well a solution.\ We may therefore form combinations of definite parity
which, up to a multiplicative constant, for $0<\eta <\alpha $ are, 
\begin{equation}
\Psi _{\pm }(\eta )\sim \frac{(\eta -\alpha )^{n}}{(\eta +\alpha )^{n+1}}%
e^{\alpha ^{2}\eta -\frac{1}{3}\eta ^{3}}\mp \frac{(\eta +\alpha )^{n}}{%
	(\eta -\alpha )^{n+1}}e^{-\alpha ^{2}\eta +\frac{1}{3}\eta ^{3}},\text{ \ \ }
\end{equation}%
Transforming to the $z$ variable, $z=2\sqrt{\alpha }(\eta -\alpha )$ and
taking the large $\alpha $ limit gives for $z>-2\alpha ^{3/2}$, 
\begin{equation}
\Psi _{\pm }(z)\rightarrow N_{\pm }\left[ z^{n}e^{-\frac{z^{2}}{4}}\mp e^{-%
	\frac{4}{3}\alpha ^{3}}2^{4n+2}\alpha ^{3(n+\frac{1}{2})}\frac{e^{\frac{z^{2}%
		}{4}}}{z^{n+1}}\right] ,\text{ \ }.  \label{s1}
\end{equation}%
Comparing Eq.\ref{s1} with Eq.\ref{Left}, we see they can be matched
provided,\bigskip 
\begin{eqnarray}
\delta _{n}(\alpha ) &=&\frac{2^{4n+\frac{3}{2}}\alpha ^{3(n+\frac{1}{2})}}{\sqrt{\pi 
	}(n+\frac{1}{2})n!}e^{-\frac{4}{3}\alpha ^{3}}  \label{Norm} \\
N_{\pm } &=&N_{n\pm }
\end{eqnarray}%
Quite remarkably the energy splittings obtained here are precisely those
obtained in the one-instanton approximation. The extremely rapid decrease of 
$e^{-\frac{4}{3}\alpha ^{3}}$\ guarantees\ that $\delta _{n}(\alpha )$ will
approach the true value for large enough $\alpha .$ In Fig.2 we compare the
leading order result obtained above with the corrections shown in Eq.\ref{m1}%
-Eq.\ref{m2}. 

\begin{figure}[h]
	\includegraphics[width=9cm, height=7cm]{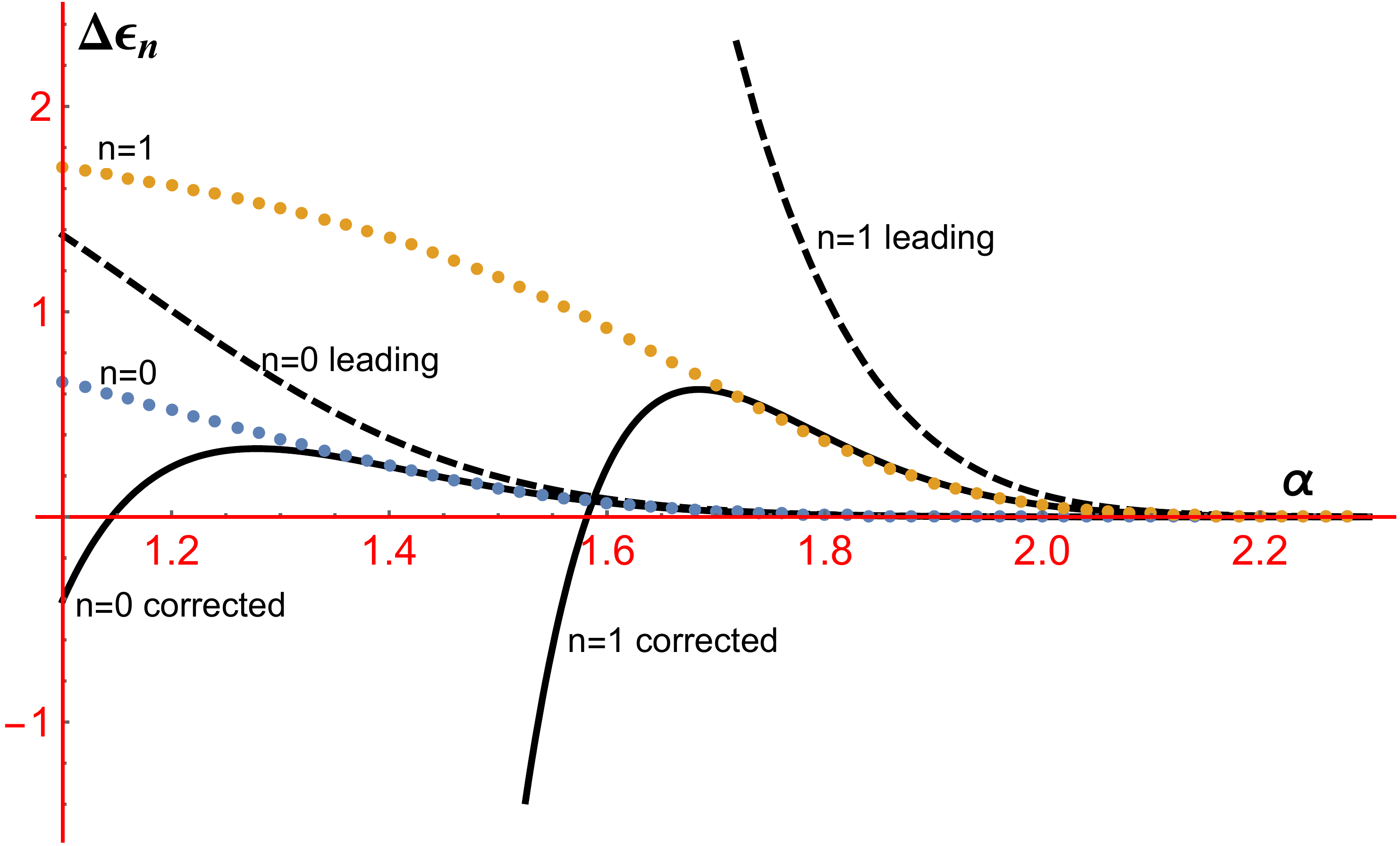} \centering
	\caption{The energy difference between odd and even states as a function of $\protect\alpha $\ for $n=0,1$. Results obtained by numerical diagonalization (dotted) are compared against the leading order result as well as with those with multi-instanton corrections.}
\end{figure}

\bigskip

\bigskip 

At this point we need to be more explicit about the domains in which
different pieces of the wavefunction are valid. The wavefunction $D_{\varepsilon -\frac{1}{2}}(z)$ , in Eq.\ref{Left} is valid around $z=0$ but not around $\eta =0$. On the other hand $\Psi _{\pm }\left(z\right) $ in Eq.\ref{s1} has the correct behavior around $\eta =0$ but fails progressively as one approaches the well bottom. At which point should they be matched? There is clearly some
ambiguity here but an obvious choice is the classical turning point to the
left of $\alpha , $ i.e. when the potential energy in Eq.\ref{q1} becomes
equal to the eigenvalues, $2\alpha ^{2}(\eta -\alpha )^{2}=\left( n+\frac{1}{
	2}\right) 2\alpha .$ This gives the two turning points in the right plane,%
\begin{equation*}
z_{\pm }=\pm \sqrt{2(2n+1)}.
\end{equation*}%
The final form of the wavefunction is,%
\begin{equation*}
\varphi _{n,\pm }(z) = \left\{ 
\begin{array}{lr}
\Psi _{\pm }(z) : & -2\alpha ^{3/2}<z<-\sqrt{2(2n+1)} \\ 
\Phi _{\pm }(z) : & -\sqrt{2(2n+1)}<z<\infty%
\end{array}
\right.
\end{equation*}
where $\Psi _{\pm }(z),\Phi _{\pm }(z)$ are defined in Eq.\ref{s1}, Eq.\ref%
{Right} respectively.

\section{Tunneling Current}

The important observable is the electron current $J_{x}$, or equivalently,
the velocity v$_{x}$. The corresponding operator is obtained from the
canonical momentum,

\begin{eqnarray}
\text{\^{v}}_{x} &=&\frac{\mathbf{p}_{x}-\frac{e}{c}\mathbf{A}_{x}}{m}\, 
\notag \\
&=&\frac{\hslash }{mL_{M}}(\eta ^{2}-\alpha ^{2})=\frac{\hslash }{mL_{M}}%
\left( \frac{z^{2}}{4\sqrt{\alpha }}+z\sqrt{\alpha }\right) .  \label{cur}
\end{eqnarray}%
In the limit of large well separations $\alpha >>\left( n+\frac{1}{2}\right)
^{1/3}$ the leading parity dependent contributions to the transverse current
of an electron state centred at $\alpha $ is, 
\begin{eqnarray}
j_{x}^{\pm } &=&e\left\langle n,\alpha ,\pm \left\vert \text{\^{v}}%
_{x}\right\vert n,\alpha ,\pm \right\rangle   \notag \\
&=&\frac{e\hslash }{mL_{M}}\left[ J_{n,\alpha }\mp \sqrt{\frac{9\pi }{2}}%
n!(n+\frac{1}{2})\alpha ^{2}\delta _{n}(\alpha )\right] .  \label{J1}
\end{eqnarray}%
Here $J_{n,\alpha }$\ is the part of the current that is parity independent.
While it is easy to get expressions for it, they cannot be evaluated
analytically and integrations will have to be done numerically. Note that if both parity states are occupied then the oppositely directed currents cancel and so, if the current is to be detected, the Fermi level must be tuned (perhaps by varying an external electric potential) so that the last positive parity occupied state lies just below its negative parity partner. The y-directed current is obviously zero when evaluated in a parity eigenstate, $\left\langle \pm \left\vert \text{\^{v}}_{y}\right\vert \pm \right\rangle =0$.

The part of the current in Eq.\ref{J1} that is proportional to $\delta _{n}(\alpha )$\ owes entirely to $\Psi_{\pm }(z).$ \ The contribution of $\Phi _{\pm }(z)$\ to this part is ignorable because it is of order $\alpha ^{0}.$\ It is easy to see why: continuity of $D_{\varepsilon -\frac{1}{2}}$ as a function of $\varepsilon $ means that any definite integral of the type $\int z^{\beta }\left\vert D_{\varepsilon -\frac{1}{2}}(z)\right\vert ^{2}dz$ that is convergent
changes linearly with $\varepsilon $, and hence linearly with $\delta
_{n}(\alpha )$ as well. On the other hand, as can be seen in Eq.\ref{J1} the
dominant contribution to the parity sensitive terms is of order $\alpha
^{2}\delta _{n}(\alpha ).$ We see that the ambiguity in determining the
matching point of the wavefunctions does not affect the leading order
result. Since $\ e^{-\frac{4}{3}\alpha ^{3}}$  decreases extremely fast with 
$\alpha $, we are essentially looking at the quantum properties of states
that lie close to the line of zero $B$ field. 

Let us return to physical considerations: an electron that is mostly on the
right ($y>0$) is in a mixture of parity states $\left[ \varphi _{+}(\eta
)+\varphi _{-}(\eta )\right] /\sqrt{2}$ whereas an electron on the left has
wavefunction $\left[ \varphi _{+}(\eta )-\varphi _{-}(\eta )\right] /\sqrt{2}%
.$ If an electron swaps its position from right to left, the net change in
the current it carries is easily seen to be, 
\begin{eqnarray}
\Delta j &=&\frac{e\hslash }{mL_{M}}\sqrt{\frac{9\pi }{2}}(n+\frac{1}{2}%
)n!\alpha ^{2}\delta _{n}(\alpha )  \label{current} \\
&=&\frac{e\hslash }{mL_{M}}6\alpha ^{\frac{7}{2}}e^{-\frac{4}{3}\alpha ^{3}}%
\text{ \ (for }n=0\text{).}\bigskip 
\end{eqnarray}%
Consider an electron initially on the right that, for example, is in the $n=0
$ state. It will will return to its original state after a time $\tau =2\pi
\hslash /\Delta E$ where $\Delta E$ is the energy splitting $\frac{\hslash
	^{2}}{mL_{M}^{2}}2\alpha \delta _{0}(\alpha ).$ Of course, no net current
will have flowed since the going and returning current will have canceled
over a complete cycle. How then are we to actually probe such states using
the expession in Eq.\ref{current}? This is taken up in the next section.

\section{Time Dependent Field}

The symmetric potential $V(\eta )$ in Eq. \ref{Ham} has zeros at $\eta =\pm
\alpha .$\ Imagine introducing a small time-dependent symmetry breaking term
proportional to $\eta $ or $\eta ^{3}$, i.e. somehow rocking this potential
so that one minimum sometimes lies slightly below the other. If this rocking
frequency can be made to coincide with the natural time for a particle to
tunnel across the potential and back, then we can expect an enhancement for
electrons at a certain distance from the $B=0$ line, i.e. those at a
particular value of $\alpha $. One way of achieving this might be to apply a
weak oscillating electric field in the $\hat{y}$ direction$.$ To this end,
let us add a perturbing term to the Hamiltonian Eq.\ref{Ham}. The time
dependent Schrodinger equation is, 
\begin{equation}
\left( -\frac{1}{2}\frac{\partial ^{2}}{\partial \eta ^{2}}+V(\eta )-\eta
a_{E}\cos \omega _{E}\tau \right) \Phi =i\frac{\partial }{\partial \tau }\Phi
\label{TD}
\end{equation}%
The dimensionless time is
$\tau =\omega _{B}t$ where, 
\begin{equation}
\omega _{B}=\frac{\hslash }{mL_{M}^{2}}
\end{equation}%
and $\omega _{E}$ is the applied frequency in units of $\omega _{B}$. The
dimensionless amplitude of the applied field is, 
\begin{equation}
\text{\ }a_{E}=\frac{2mc}{\hslash B^{\prime }}E_{y}.
\end{equation}%
If we choose only small amplitudes $a_{E}$ there will be negligible mixing
with higher $n$ levels in which case it should be sufficient to confine
ourselves to a two-dimensional space,%
\begin{equation}
\Phi =\beta _{+}(\tau )e^{-i\varepsilon _{+}\tau }\varphi _{n+}(\eta )+\beta
_{-}(\tau )e^{-i\varepsilon _{-}\tau }\varphi _{n-}(\eta ).  \label{TDS}
\end{equation}%
Although the eigenfunctions and eigenvalues have been derived for arbitrary $%
n$, only the lowest values of $n$ are likely to be relevant. To avoid
notational clutter, let us therefore specialize to $n=0$. The equations for
the amplitudes $\beta _{\pm }$ follow straightforwardly from Eq.\ref{TD},%
\begin{eqnarray}
i\text{ }\dot{\beta}_{+} &=&-a_{E}\alpha \beta _{-}e^{-i\omega _{D}\tau
}\cos \omega _{E}\tau \label{k1} \\
i\text{ }\dot{\beta}_{-} &=&-a_{E}\alpha \beta _{+}e^{i\omega _{D}\tau }\cos
\omega _{E}\tau \label{k2}
\end{eqnarray}%
Here the difference frequency $\omega _{D}$ is, 
\begin{eqnarray}
\omega _{D} &=&\varepsilon _{-}-\varepsilon _{+} \\
&=&8\sqrt{\frac{2\alpha ^{5}}{\pi }}e^{-\frac{4\alpha ^{3}}{3}}.
\end{eqnarray}

The coupled equations are essentially those of the well-known Rabi
oscillations. Eqs.\ref{k1}-\ref{k2} can be solved in the well known rotating wave approximation \cite{Fujii}. This amounts to replacing $e^{-i\omega _{0}\tau
}\cos \omega _{E}\tau $ with $\frac{1}{2}e^{-i(\omega _{0}-\omega _{E})\tau
} $ and $e^{i\omega _{0}\tau }\cos \omega _{E}\tau $ with $\frac{1}{2}%
e^{i(\omega _{0}-\omega _{E})\tau }$ (this, in effect, throws away the rapid
oscillation leaving only the slowly modulated envelope). If the system was
initially in the positive parity ground state at $\tau =0$\ the subsequent
probabilities are,%
\begin{eqnarray}
\left\vert \beta _{+}\right\vert ^{2} &=&1-4\frac{\alpha ^{2}a_{E}^{2}}{
	\Omega ^{2}}\sin ^{2}\frac{\Omega \tau }{2} \\
\left\vert \beta _{-}\right\vert ^{2} &=&4\frac{\alpha ^{2}a_{E}^{2}}{\Omega
	^{2}}\sin ^{2}\frac{\Omega \tau }{2} \\
\Omega ^{2} &=&4\alpha ^{2}a_{E}^{2}+\left( \omega _{D}-\omega _{E}\right)
^{2}.
\end{eqnarray}

The current operator in Eq.\ref{TDS}. can be inserted into the state $\Phi $
in Eq.\ref{TDS}. Now the current has an AC component proportional to $\cos
\Omega \tau $,\bigskip 
\begin{equation}
\Delta j(\alpha )=3\frac{e\hslash }{mL_{M}}\alpha ^{\frac{7}{2}}e^{-\frac{4}{%
		3}\alpha ^{3}}\frac{\alpha ^{2}a_{E}^{2}}{\alpha ^{2}a_{E}^{2}+\frac{1}{4}%
	\left( \omega _{D}(\alpha )-\omega _{E}\right) ^{2}}\cos \Omega \tau .
\end{equation}%
This is current of a single oscillator and the contributions of all
oscillators must be added up. This is easily done using Eq.\ref{dens},

\begin{equation}
\Delta J=\frac{2}{\pi }\frac{L_{x}}{L_{M}}\int_{0}^{\alpha _{F}}d\alpha
\alpha \Delta j(\alpha ).
\end{equation}%
The factor of 2 has been inserted to account for the two electron spin
directions - the electron levels will be split by the Zeeman splitting but
this has been neglected in this initial investigation. The upper limit of
the integral is determined by the Fermi momentum $k_{F},$ $\alpha _{F}=\sqrt{
	k_{F}\L _{M}}.$ Note that only the region of positive $k_{x}$ contributes, i.e. for $\alpha ^{2}>0$. This is evident from the spectrum (see Fig.1) because opposite parity states for  $\alpha ^{2}<0$ are split by a huge difference. After some simplification,

\begin{equation}
\Delta J=\frac{6}{\pi }\frac{e\hslash L_{x}}{mL_{M}^{2}}\int_{0}^{\alpha
	_{F}}d\alpha \frac{a_{E}^{2}\alpha ^{\frac{9}{2}}e^{-\frac{4}{3}\alpha ^{3}}%
}{\alpha ^{2}a_{E}^{2}+\frac{1}{4}\left( \omega _{D}(\alpha )-\omega
	_{E}\right) ^{2}}\cos \Omega \tau .  \label{delJ}
\end{equation}
This result, specialized to the $n=0$ case, can be generalized to arbitrary $%
n$. Larger values of $n$ will, however, lead to less tunneling. The integral
above receives contributions from all values of $\alpha $, both large and
small but our analysis is only asymptotically valid for large $\alpha .$
Fortunately one may take advantage of the resonance factor which spikes at $%
\omega _{D}(\alpha _{0})=\omega _{E}$,%
\begin{equation}
R(\alpha )=\frac{1}{1+\frac{\omega _{D}^{\prime }(\alpha _{0})^{2}}{4\alpha
		_{0}^{2}a_{E}^{2}}\left( \alpha -\alpha _{0}\right) ^{2}}.
\end{equation}%
For illustrative purposes we take the limit of an infinitely weak external
field $a_{E}\rightarrow 0$. Using,

\begin{equation*}
\int_{-\infty }^{+\infty }d\alpha R(\alpha )=\pi \frac{2\alpha _{0}a_{E}}{
	\left\vert \omega _{D}^{\prime }(\alpha _{0})\right\vert },
\end{equation*}%
allows us to approximate the integral in Eq.\ref{delJ} for \ $\alpha
_{0}<\alpha _{F}$,%
\begin{equation}
\Delta J_{res}=12\frac{e\hslash L_{x}}{mL_{M}^{2}}\alpha _{0}^{\frac{9}{2}%
}e^{-\frac{4}{3}\alpha _{0}^{3}}\frac{1}{\left\vert \omega _{D}^{\prime
	}(\alpha _{0})\right\vert }a_{E}\cos 2\alpha _{0}a_{E}\tau .\text{ \ }
\end{equation}%
By choosing the externally applied field's frequency appropriately, in
principle this allows us to selectively stimulate the tunneling of only
those electrons sufficiently far away from the $B_{z}=0$ line where the
asymptotically correct expressions obtained in this paper are valid. Of
course, in practice a finite range of $\alpha $\ values will contribute and\
one would need to do the integral in Eq.\ref{delJ} numerically. That $\Delta
J_{res}$\ increases linearly with the length of the sample is readily
understood - the number of oscillators with frequency $\omega _{D}$ is
proportional to the length of the $B_{z}=0$ line, i.e. one is effectively
dealing with a quantum wire.\ 

\section{Discussion}
This paper has focussed upon electrons close to the $B_{z}=0$ line in a
2DEG. With a suitable choice of vector potential, and after making
appropriate choices of variables, this reduces to the study of the famous
double well potential as in Eq.\ref{SE1}. This does not have exact
solutions. In the process of seeking approximate solutions we derived the
splitting between opposite parity states. This result coincides with that
for a dilute gas of instantons.\ The instanton method, although it lends
itself to systematic improvement, is considerably more involved. However
this method has not, to our knowledge, yet been applied to the calculation
of the currents or observables other than energy. We therefore directly used
the Schrodinger equation to find the associated wavefunctions from which one
may compute the expectation values of any single particle operator.

The preliminary investigation reported here must face challenges in both the
theoretical and experimental domains. Let us start with reviewing the
assumptions of the former. We have constructed a wavefunction by insisting
upon boundary conditions at $y=0$ with no external confinement (i.e. without
imposing $\varphi (\pm \frac{L_y}{2})=0$ at the edges). This is well justified
because our interest is only on electron behaviour along the $B_{z}=0$ line
rather than in the bulk or the edges (where there would be skipping orbitals
in the case of a hard wall). The philosophy used here was that the large $%
\alpha $ limit would yield asymptotically correct energy splittings and
wavefunctions. Indeed, for energies this can be explicitly verified (see
Fig. 2). But wavefunctions - particularly for excited states - are more
difficult to deal with. The philosophy used here was to match parabolic
cylinder functions that are valid inside the wells with those that have the
correct symmetries and are correct in the large $\alpha $\ limit close to $%
y=0$. Inevitably there arises the question of where to match. Fortunately,
the leading order contribution to $\left\langle j_{x}\right\rangle $\ is
independent of the matching point. But this does not mean, of course, that
subleading terms will be similarly independent. Presumably variational
wavefunctions can be constructed that are valid in the entire space and
which also embody the correct symmetries. One would then know better where
the asymptotic region sets in.

Let us now look at experimental prospects. Current technology using cobalt
magnets allows the making of 2DEGs with strong magnetic field gradients of
around $1G/\mathring{A}$ which corresponds to a magnetic length scale $%
L_{M}\sim 1.1\times 10^{3}\mathring{A}.$ \ For purposes of studying effects
close to the $B_{z}=0$ line, the boundary effects can be made negligible by
taking the sample width $L_{y}$ to be greater than $10^{4}\mathring{A}$ .
This is because the of the extremely rapid decay factor $e^{-\frac{4}{3}
	\alpha ^{3}}.$ Reducing the effects of impurities would be crucial. If
translational symmetry in the $x$-direction is broken then $k_{x}$ would not
remain a good quantum number and so one would have non-diagonal transitions, 
$\left\langle \alpha ^{\prime }\left\vert \text{\^{v}}_{x}\right\vert \alpha
\right\rangle \neq 0$. Fortunately high purity samples exist - Hara et al 
\cite{Expt1} report using a 2DEG with electron mean free path $\lambda $ of
around 6100 $\mathring{A}$ which is roughly 6 times larger than $L_{M}.$
This is a hopeful sign although a proper calculation is needed to see how
large $\lambda $\ must be for impurity scattering to play a negligible part.
Perhaps the greater challenge would be to achieve ultra-low electron
densities. In principle tunneling can occur from any partially filled $n$
state. But, as has been repeatedly emphasized, calculations in the $n=0$
band are easier and more reliable than for higher $n$'s because of the
relative simplicity of the wavefunctions. Also, tunneling effects decrease
with increasing $n$.

To conclude: it appears - at least in principle - that it is possible to
experimentally investigate the mechanism by which spatially separated
electrons gyrating in Landau orbits can cross over to their equilibrium
positions on the other side of the potential barrier through tunneling. At
the calculational level the preliminary investigation reported here is
accurate only for relatively large well separations and more precise
calculations would be needed for smaller $\alpha $. Therefore we have here an
interesting laboratory for doing instanton calculations beyond the dilute gas
approximation or for higher order WKB approximations. This simple physical system allows for some other investigations as well. For ordinary wavepacket tunneling, Davies \cite{Davies} has discussed the subtleties involved in calculating tunneling time, and suggested the use of a "quantum clock" attached to the particle. In the present context one can again explore how much time is needed to actually transmit information across the wire (rather than along it), i.e. for a wavepacket to make it to the other side.

\bigskip

\bigskip 
{\Large Acknowledgments\medskip }

The author thanks Amer Iqbal, A.H. Nayyar, Kashif Sabeeh, and Sohail Zubairy
for discussions and comments. He is especially grateful to Shivaji Sondhi
for pointing out key references and for a discussion on experimental
possibilities.


\begin{thebibliography}{99}
	\bibitem{Muller} Effect of a non-uniform magnetic field on a two-dimensional
	electron gas in the ballistic region, J.E.Muller, Phys.Rev.Lett. 68, p385
	(1992).
	
	\bibitem{Reijners1} Snake orbits and related magnetic edge states, J.
	Reijniers and F. M. Peeters, J. Phys.: Cond. Matter \textbf{12} 9771--9786
	(2000).
	
	\bibitem{Reijners2} Confined magnetic guiding orbit states, Reijniers J,
	Matulis A, Chang K, Peeters F M and Vasilopoulos P, Europhys. Lett. \textbf{%
		\ 59} 749 (2002).
	
	\bibitem{Nogaret} Electron dynamics in inhomogeneous magnetic fields, Alain
	Nogaret, J. Phys, Cond. Matter \textbf{22}, 253201, (2010).
	
	\bibitem{Rectification} Electrical rectification by magnetic edge states,
	Lawton D, Nogaret A, Makarenko M V, Kibis O V, Bending S J and Henini M,
	Physica E \textbf{13} 699 (2002).
	
	\bibitem{Expt2} Observation of quantum states without a semiclassical
	equivalence bound by a magnetic field gradient, B. Schuler, M. Cerchez,
	Hengyi Xu, J. Schluck, T. Heinzel, D. Reuter, and A. D. Wieck, Phys. Rev. B 
	\textbf{90}, 201111(R) (2014).
	
	\bibitem{Expt1} Transport in a two-dimensional electron-gas narrow channel
	with a magnetic-field gradient, Hara M, Endo A, Katsumoto S and Iye Y 2004,
	Phys. Rev. B \textbf{69} 153304.
	
	\bibitem{Coleman} Aspects of Symmetry - Selected Lectures of Sidney Coleman,
	Cambridge University Press , page 265 (1985).
	
	\bibitem{Fujii} Introduction to the Rotating Wave Approximation (RWA) : Two
	Coherent Oscillations, Kazuyuki Fujii, arXiv:1301.3585v3.
	
	\bibitem{Zinn1} Multi-instantons and exact results I: Conjectures, WKB
	expansions, and instanton interactions, J.Zinn-Justin and U.D.Jentschura,
	Annals Phys. \textbf{313}, 197 (2004), \qquad arXiv:quant-ph/0501136.
	
	\bibitem{Zinn2} J. Zinn-Justin and U. D. Jentschura, Multi-instantons and
	exact results II: Specific cases, higher-order effects, and numerical
	calculations," Annals Phys. \textbf{313}, 269 (2004), arXiv:quant-ph/0501137.
	
	\bibitem{Bender1} Anharmonic Oscillator, C. M. Bender and T. T. Wu, Phys.
	Rev. \textbf{184}, 1231 (1969).
	
	\bibitem{Bender2} Anharmonic Oscillator 2: A Study of Perturbation Theory in
	Large Order, C. M. Bender and T. T. Wu, Phys. Rev. D 7, 1620, 43, (1973).
	
	\bibitem{Davies} Quantum tunneling time, P.C.W. Davies, Am.J.Phy. \textbf{73}, 23 (2005) arXiv:quant-ph/0403010.
\end{thebibliography}
\end{document}